\documentclass[11pt,a4paper]{article}
\usepackage{jheppub}

\usepackage[T1]{fontenc}
\usepackage{amssymb,amsmath}
\usepackage{lmodern}
\usepackage{verbatim,setspace}

\newcommand{\Scal}[1]{\Bigl ({#1} \Bigr )}
\newcommand{\scal}[1]{\bigl ({#1} \bigr )}
\def\ie{{\it i.e.}\ }

\DeclareMathAlphabet{\mathpzc}{OT1}{pzc}{m}{it}
\usepackage{dsfont}
\usepackage{amssymb}
\usepackage{amsmath}

\newcommand{\ord}[1]{{\scriptscriptstyle (#1)}}

\def\cN{{\mathcal{N}}}
\def\gl{\mathfrak{gl}}

\newcommand{\Pp}{\Phi_{\scriptscriptstyle +}}
\newcommand{\Pm}{\Phi_{\scriptscriptstyle -}}
\newcommand{\Ppm}{\Phi_{\scriptscriptstyle \pm}}
\newcommand{\Pmp}{\Phi_{\scriptscriptstyle \mp}}

\renewcommand\Re{\hbox{{\rm Re}}\,}

\def\bea{\begin{eqnarray}}
\def\eea{\end{eqnarray}}
\def\be{\begin{equation}}
\def\ee{\end{equation}}

\newcommand{\CR}{\nonumber \\*}

\def\cH{{\mathcal H}}
\def\cV{{\mathcal V}}

\newcommand{\pA}{{\text{\tiny A}}}

\def\cE{{\mathcal{E}}}

\usepackage{color}

\makeatletter
\newcommand{\thickhline}{%
    \noalign {\ifnum 0=`}\fi \hrule height 1.3pt
    \futurelet \reserved@a \@xhline
}

\title{Floating JMaRT}
\author[a]{Guillaume Bossard,}  \author[b]{and Stefanos Katmadas}
\affiliation[a]{Centre de Physique Th\'eorique, Ecole Polytechnique, CNRS, 91128
Palaiseau, France}
\affiliation[b]{Dipartimento di Fisica, Universit\'a di Milano-Bicocca and INFN, sezione di Milano--Bicocca,\\
Milano, Italy}

\emailAdd{guillaume.bossard [at] cpht.polytechnique.fr} \emailAdd{stefanos.katmadas [at] unimib.it}

\abstract{
We define a new partially solvable system of equations that parametrises solutions to six-dimensional $\cN=(1,0)$ ungauged
supergravity coupled to tensor multiplets. We obtain this system by applying a series of dualities on
the known floating brane system, imposing that it allows for the JMaRT solution. We construct an explicit
multi-centre solution generalising the JMaRT solution, with an arbitrary number of additional BPS centres on a line. We describe explicitly the embedding of the JMaRT solution in this system in five dimensions. 
}

\preprint{CPHT-RR096.1214}

\keywords{Supergravity theories, Black holes in string theory}
\begin{document}
 \maketitle

\section{Introduction}

The investigation of smooth solitonic solutions to supergravity has attracted considerable interest in the last
years, due to the fuzzball proposal of Mathur and collaborators \cite{Mathur:2005ai,Mathur:2005zp}. According to
this proposal, a black hole quantum state is not well approximated by its classical geometry near its horizon,
but is rather defined as a sum of microstates that are themselves well approximated by globally hyperbolic smooth
geometries. While the search for the possible smooth geometries has met with some success in the case of extremal black holes,
especially in the presence of supersymmetry 
\cite{Mathur:2003hj,Bena:2005va,Berglund:2005vb,Saxena:2005uk,Bena:2007kg,Bena:2010gg,Lunin:2012gp, Bobev:2011kk,Niehoff:2013mla},
less progress has been made for the non-extremal black holes, for which only a handful of special solutions are known
\cite{Jejjala:2005yu,Bena:2009qv,Bobev:2009kn,Giusto:2007tt,Compere:2009iy,Banerjee:2014hza}.

An important step was made by the introduction of the floating brane ansatz
\cite{Bena:2009qv,Bobev:2009kn,Bena:2009fi,Bossard:2014yta},
which led to a partially solvable system of equations describing non-extremal solutions, so that a systematic search for
smooth solutions is simplified. This is a system of equations that is based on an Euclidean Maxwell--Einstein subsystem,
which governs the spatial section of the metric and generalises the hyper--K\"{a}hler metrics appearing in BPS solutions.  
However, it has been noted in \cite{Gibbons:2013tqa, Bossard:2014yta} that solutions
to this system generically lack a spin structure and therefore seem unattractive as microstate geometries. This is
not so for the JMaRT solution of \cite{Jejjala:2005yu}, which is not described by the floating brane ansatz and is
known to have many desirable features for a microstate geometry, including a dual CFT description, despite the fact
that it violates the over-rotating regularity bound for the associated asymptotically Minkowski black hole solution in five dimensions.

In this paper, we define a new partially solvable system describing solutions to six-dimensional $\cN=(1,0)$ supergravity,
which we refer to as the floating JMaRT system. This system includes the JMaRT solution by construction and may therefore
lead to more physically interesting multi-centre solutions. To obtain this system we apply well chosen three-dimensional duality transformations to the floating brane ansatz based on a Maxwell--Einstein instanton, and determine the parameters of the duality such that the new system includes the JMaRT solution. The result is an inequivalent ansatz for the supergravity solution, defined in terms of the same functions appearing in the floating brane system for supergravity in five dimensions. Even though the JMaRT solution is defined in six dimensions and the system we define in this paper a priori only admits smooth globally hyperbolic solutions in six dimensions, we write the explicit ansatz in five dimensions, and only briefly discuss its uplift to six dimensions. This simplifies to some extent the comparison with the JMaRT solution in the simplified form given in \cite{Gimon:2007ps}.

Since the new system is again based on a Maxwell--Einstein subsystem, one may obtain solutions by applying exactly the same methods as for the floating brane ansatz. However, the Euclidean four-dimensional metric is not any more realised geometrically in supergravity. This is necessary, since the Euclidean base of the JMaRT solution admits a non-trivial Ricci scalar, whereas the Maxwell energy momentum tensor is traceless. Despite the fact that the system of equations is similar to the one appearing in the floating brane system, the explicit ansatz for the solution turns out to be considerably more complicated.

The structure of the paper is organised as follows: In section \ref{sec:JM-sys} we discuss the manipulations required to obtain
the floating JMaRT system, giving its explicit form and the ansatz for the various fields in five dimensions. We then go on in section
\ref{sec:ex-sols} to discuss the generalisation of the multi-centre Maxwell--Einstein instantons of \cite{Bossard:2014yta} to include angular momentum at the bolt. We then solve the floating JMaRT system based on these instantons, giving the explicit form of the functions appearing in the ansatz for the general case. We then specialise to
the case with a single non-extremal centre and show that the JMaRT solution is indeed contained in this class by
appropriately fixing some of the parameters. We conclude with some remarks on future directions in \ref{sec:concl}.
Finally, in appendix \ref{eq:4d-sys} we briefly discuss the floating JMaRT system in four dimensions.

\section{The floating JMaRT system}
\label{sec:JM-sys}
In this section we describe how to obtain a partially solvable system of differential equations which allows for the JMaRT solution. The method we use is based on the three-dimensional non-linear sigma model over a para-quaternionic symmetric space one obtains after dimensional reduction of $\cN=(1,0)$ supergravity in six dimensions along two space-like and one time-like isometries. After a brief discussion of the dimensional reduction down to three dimensions, we discuss the specific three-dimensional duality transformation one has to apply to the partially solvable system defined in \cite{Bena:2009fi} within the floating brane ansatz, in order to obtain the floating JMaRT system.

\subsection{Euclidean Rasheed--Larsen duality}
The JMaRT solution is a smooth solution of $\cN=(1,0)$ supergravity in six dimensions coupled to one tensor multiplet. In this section we will review how to obtain partially solvable systems solving supergravity equations of motion \cite{Bossard:2014yta}, and will construct a particular system that includes the JMaRT solution \cite{Jejjala:2005yu}. The procedure applies to any (ungauged) $\cN=(1,0)$ supergravity in the Jordan family \cite{Gunaydin:2010fi}, but for simplicity we will consider the example of the theory associated to the Jordan algebra of real three by three symmetric matrices, which includes 2 vector multiplets and 2 tensor multiplets in six dimensions. We assume that the six-dimensional space-time admits three commuting isometries of respective Killing vectors $\partial_y,\, \partial_\psi,\, \partial_t$, such that the equations of motions effectively reduce to the ones of a non-linear sigma model over a  para-quaternionic symmetric space coupled to Euclidean gravity in three dimensions. The scalar fields parametrise the symmetric spaces \cite{Gunaydin:1983bi}
\begin{equation}
 SL(2)/SO(2) \underset{\partial_y}{\longrightarrow}  SL(3)/SO(3)   \underset{\partial_\psi}{\longrightarrow} Sp(6,\mathds{R})/U(3)  \underset{\partial_t}{\longrightarrow} F_{4(4)}/( SL(2)\times_{\mathds{Z}_2} Sp(6,\mathds{R})) 
\end{equation}
in 6, 5, 4 and 3 dimensions respectively, where the Killing vectors under the arrows label the coordinates eliminated by dimensional reduction in each step. The fields of the six-dimensional theory are then determined by the scalar fields parametrising the para-quaternionic space $F_{4(4)}/( SL(2)\times_{\mathds{Z}_2} Sp(6,\mathds{R})) $ in the parabolic gauge associated to the graded decomposition 
\begin{multline} 
\mathfrak{f}_{4(4)}  \cong   \overline{\bf 3}^\ord{-4}\oplus {\bf 2}^\ord{-3}\oplus({\bf 3}\otimes 
{\bf 3})^\ord{-2}  \oplus (\overline{\bf 3}\otimes {\bf 2})^\ord{-1}\\ \oplus  \scal{ \gl_1  \oplus 
\mathfrak{sl}_3\oplus  \mathfrak{sl}_2 }^\ord{0}  \oplus ({\bf 3}\otimes {\bf 2})^\ord{1} \oplus(\overline{\bf 
3}\otimes {\bf 3})^\ord{2}\oplus {\bf 2}^\ord{3} \oplus  {\bf 3}^\ord{4}\ ,\label{ParaGrad}  
\end{multline}
where the odd positive grade components correspond to the 2 vector fields and the even positive grade components to the 3 chiral tensor fields. The equations of motion for the coset representative, $\cV$, in $F_{4(4)}/( SL(2)\times_{\mathds{Z}_2} Sp(6,\mathds{R})) $ is defined from the component $P$ of the Maurer--Cartan form 
\begin{equation}\cV^{-1} d\cV = P + B  \  \end{equation}
in the coset component $\mathfrak{f}_{4(4)} \ominus (\mathfrak{sl}_2 \oplus \mathfrak{sp}(6,\mathds{R}))$. The equations of motion can then be cast as the conservation law for the $\mathfrak{f}_{4(4)}$ current and the three-dimensional Einstein equation, as
\begin{equation}\label{eq:3Deoms}
d \star \scal{ \cV P \cV^{-1} }= 0 \ , \qquad 
R_{ij} = \frac{1}{12} \mbox{Tr} \,   P_i P_j \  .
\end{equation}
There is an inequivalent conjugacy class of embeddings of the parabolic subgroup of $F_{4(4)}$ defined from \eqref{ParaGrad}, such that the $GL(1)$ subgroup lies inside the divisor subgroup $SL(2)\times  Sp(6,\mathds{R})$. Such group embedding defines by construction a consistent truncation of the theory through the coset space embedding
\begin{equation}
 \scal{ SL(3)\ltimes \mathds{R}^{3\times2+3\times3+2+3}}\big/\scal{GL(2)\ltimes \mathds{R}^{2\times 2 + 3+0+1}} \subset F_{4(4)}/( SL(2)\times_{\mathds{Z}_2} Sp(6,\mathds{R})) \ . \label{ParaEmbed}
\end{equation}
We shall consider the partially solvable system defined by constraining the coset representative $\cV$ to lie in this parabolic subgroup. In this case one straightforwardly checks that only the component parametrising $SL(3)/GL(2)$ sources the Euclidean Riemann tensor in \eqref{eq:3Deoms}, and the conservation of the current preserves the graded decomposition \eqref{ParaGrad}, so that once the equations of motion for the $SL(3)/GL(2)$ non-linear sigma model coupled to Euclidean gravity are solved, the twelve remaining fields satisfy to a solvable linear system of equations. 

The $SL(3)/GL(2)$ non-linear sigma model coupled to Euclidean gravity describes solutions to the four-dimensional Euclidean Maxwell--Einstein theory admitting a $U(1)$ isometry. Therefore, the starting point to obtain a solution to the system described above is to choose an appropriate Maxwell--Einstein instanton, which must not necessarily be a regular solution by itself. We will describe a particular family of instantons in section \ref{sec:ein-max}. Because such solutions define by construction a Riemannian four-dimensional metric, it is natural to consider a particular embedding \eqref{ParaEmbed}, such that this metric defines the Euclidean base in five dimensions. Such an embedding leads then to the system defined in the floating brane ansatz \cite{Bena:2009fi}, as discussed in \cite{Bossard:2014yta}. However, the JMaRT solution in five dimensions features an Euclidean base space with a non-vanishing Ricci scalar \cite{Gimon:2007ps}, and cannot be a solution to Euclidean Maxwell--Einstein equations, since the Maxwell energy momentum tensor is traceless. In order for the system to include the JMaRT solution, it is therefore necessary to consider an embedding \eqref{ParaEmbed} such that the Euclidean Maxwell--Einstein metric is not realised geometrically in the supergravity solution. 

All the embeddings \eqref{ParaEmbed} are conjugate under the action of $F_{4(4)}$, so we simply need to consider the relevant duality rotation to transform the floating brane system into a system including the JMaRT solution. For this purpose, we need an $F_{4(4)}$ transformation that can turn on the Euclidean Ricci scalar. This can be realised if this transformation rotates the Maxwell--Einstein instanton to a solution of Kaluza--Klein theory, which includes a scalar field in four dimensions to source the Ricci scalar. It is therefore rather natural to use the same type of duality transformation that was used in \cite{Rasheed:1995zv,Larsen:1999pp} to obtain Kaluza--Klein black holes. We will now describe this duality transformation in $F_{4(4)}$. 

For this purpose, it will be convenient for us to consider the parabolic gauge relevant to describe the fields of the theory in four dimensions, and we write the $F_{4(4)}$ representative in the fundamental ${\bf 26}$ representation as a constrained 27 by 27 $E_{6(6)}$ matrix that acts trivially on the antisymmetric tensor component proportional to the symplectic form, according to the conventions used in \cite{Bossard:2012ge}, \ie  
\begin{eqnarray}
\cV &=&  \exp\bigl[ \zeta^{ABC} {\bf E}_{ABC} + \varsigma {\bf E} \bigr] \exp[ U {\bf H}] \left(\begin{array}{ccc} 
\hspace{2mm} v_a{}^{A^\prime}  \hspace{2mm}& \hspace{2mm} 0  \hspace{2mm}  & \hspace{2mm}0 \hspace{2mm}  \\
\hspace{2mm}0  \hspace{2mm}& \hspace{2mm} v_a{}^{A^\prime}  \hspace{2mm}  & \hspace{2mm} 0  \hspace{2mm}  \\
\hspace{2mm}  0    \hspace{2mm}& \hspace{2mm}  0 \hspace{2mm}  & \hspace{2mm} (v^{-1})_{[B^\prime}{}^b (v^{-1})_{C^\prime]}{}^c \hspace{2mm} \end{array} \right)  \begin{array}{c} \\ \\ \vspace{4mm} \end{array} \CR
&=& \left(\begin{array}{ccc} 
\hspace{2mm} e^U v_a{}^{A}  \hspace{2mm}& \hspace{2mm} e^{-U} v_a{}^{A} \varsigma + \frac{1}{2} e^{-U} v_a{}^{B}{}  \zeta^{ADE} \zeta_{BDE}     \hspace{2mm}  & \hspace{2mm} (v^{-1})_D{}^b (v^{-1})_E{}^c  \zeta^{ADE} \hspace{2mm}  \\
\hspace{2mm}0  \hspace{2mm}& \hspace{2mm}e^{-U} v_a{}^{A}  \hspace{2mm}  & \hspace{2mm} 0  \hspace{2mm}  \\
\hspace{2mm}  0    \hspace{2mm}& \hspace{2mm}  e^{-U} v_a{}^{D}  \zeta_{BCD} \hspace{2mm}  & \hspace{2mm} (v^{-1})_{[B}{}^b (v^{-1})_{C]}{}^c \hspace{2mm} \end{array} \right)  \ . 
\end{eqnarray}
Here, $v_a{}^A$ is the $Sp(6,\mathds{R})$ matrix $Sp(6,\mathds{R})/U(3)$ coset  representative parametrised by the scalar fields, understood as the transverse $v^t$ acting on the right through the contraction with the index $a$ of the local $SO(6) \supset U(3)$, and the index $A$ of the rigid $ SL(6,\mathds{R})\supset Sp(6,\mathds{R})$. Moreover, the $\zeta_{ABC}$ determine the time components of the $14$ electro-magnetic vector fields in the rank three antisymmetric symplectic traceless representation of $Sp(6,\mathds{R})$, and $e^{2U}$ is the scaling factor of the four-dimensional metric, whereas $\varsigma$ is the scalar field dual to the vector component of the metric. We use the projective coordinates for the scalar fields, combined into a complex 3 by 3 symmetric matrix ${\bf z} = \frac{ \hspace{-1.5mm} \bf X}{X^0}$, of eigenvalues of strictly positive imaginary part \cite{Gunaydin:1983bi}, so that the pre-potential is 
\be \mathcal{F}[X] = - \frac{ \det {\bf X}}{X^0} \ .  \ee
In this basis, the Euclidean version of the duality used by Rasheed and Larsen is
\be g_E = \left(\begin{array}{ccc} 
\hspace{2mm} \delta^{A}_D  \hspace{2mm}& \hspace{2mm} 0 \hspace{2mm}  & \hspace{2mm} 0  \hspace{2mm}  \\
\hspace{2mm} k_3 \delta^A_D  \hspace{2mm}& \hspace{2mm}\delta_D^{A}  \hspace{2mm}  & \hspace{2mm} f^{AEF}  \hspace{2mm}  \\
\hspace{2mm}  f_{BCD}     \hspace{2mm}& \hspace{2mm} 0 \hspace{2mm}  & \hspace{2mm}\delta^{[E}_{[B} \delta_{C]}^{F]}  \hspace{2mm} \end{array} \right) \ , \ee
where $k_3$ is a constant and $ f^{AEF} $ is a rank one charge vector along the direction $p^0$, given by
\be f^{AEF}  \hspace{0.9mm} \hat{=} \hspace{0.9mm} \frac {1 + 6 k_ 3^{\; 2}} {2 k_ 3} \left(\begin{array}{ccccccccccccccc} 
\hspace{0.5mm} 0 \hspace{0.5mm}& \hspace{0.5mm} 1 \hspace{0.5mm}  & \hspace{0.5mm} 0 \hspace{0.5mm} & \hspace{0.5mm} 0   \hspace{0.5mm} & \hspace{0.5mm} 0   \hspace{0.5mm} & \hspace{0.5mm} 0   \hspace{0.5mm} & \hspace{0.5mm} 0 \hspace{0.5mm} & \hspace{0.5mm} 0  \hspace{0.5mm} & \hspace{0.5mm} 0 \hspace{0.5mm} & \hspace{0.5mm}0 \hspace{0.5mm} & \hspace{0.5mm}0  \hspace{0.5mm} & \hspace{0.5mm} 0 \hspace{0.5mm}& \hspace{0.5mm} 0 \hspace{0.5mm} & \hspace{0.5mm}0 \hspace{0.5mm} & \hspace{0.5mm} 0  \hspace{0.5mm} \\
\hspace{0.5mm} 0 \hspace{0.5mm}& \hspace{0.5mm} 0  \hspace{0.5mm}  & \hspace{0.5mm} 0  \hspace{0.5mm} & \hspace{0.5mm} 0   \hspace{0.5mm} & \hspace{0.5mm} 0  \hspace{0.5mm} & \hspace{0.5mm} 0  \hspace{0.5mm} & \hspace{0.5mm} 0  \hspace{0.5mm} & \hspace{0.5mm} 0  \hspace{0.5mm} & \hspace{0.5mm} 0  \hspace{0.5mm} & \hspace{0.5mm} 0 \hspace{0.5mm} & \hspace{0.5mm} 0  \hspace{0.5mm} & \hspace{0.5mm}0  \hspace{0.5mm}& \hspace{0.5mm} 0   \hspace{0.5mm} & \hspace{0.5mm} 0 \hspace{0.5mm} & \hspace{0.5mm} 0 \hspace{0.5mm} \\

\hspace{0.5mm}0  \hspace{0.5mm}& \hspace{0.5mm} 0 \hspace{0.5mm}  & \hspace{0.5mm} 0  \hspace{0.5mm} & \hspace{0.5mm} 0  \hspace{0.5mm} & \hspace{0.5mm} 0   \hspace{0.5mm} & \hspace{0.5mm} 0 \hspace{0.5mm} & \hspace{0.5mm} 1 \hspace{0.5mm} & \hspace{0.5mm} 0   \hspace{0.5mm} & \hspace{0.5mm} 0  \hspace{0.5mm} & \hspace{0.5mm} 0   \hspace{0.5mm} & \hspace{0.5mm} 0  \hspace{0.5mm} & \hspace{0.5mm} 0 \hspace{0.5mm}& \hspace{0.5mm} 0  \hspace{0.5mm} & \hspace{0.5mm} 0 \hspace{0.5mm} & \hspace{0.5mm}0  \hspace{0.5mm} \\
\hspace{0.5mm} 0  \hspace{0.5mm}& \hspace{0.5mm}0  \hspace{0.5mm}  & \hspace{0.5mm} 0   \hspace{0.5mm} & \hspace{0.5mm} 0  \hspace{0.5mm} & \hspace{0.5mm} 0   \hspace{0.5mm} & \hspace{0.5mm} 0 \hspace{0.5mm} & \hspace{0.5mm} 0  \hspace{0.5mm} & \hspace{0.5mm} 0    \hspace{0.5mm} & \hspace{0.5mm} 0  \hspace{0.5mm} & \hspace{0.5mm} 0    \hspace{0.5mm} & \hspace{0.5mm} 0  \hspace{0.5mm} & \hspace{0.5mm} 0 \hspace{0.5mm}& \hspace{0.5mm} 0  \hspace{0.5mm} & \hspace{0.5mm}0 \hspace{0.5mm} & \hspace{0.5mm} 0 \hspace{0.5mm} \\

\hspace{0.5mm} 0  \hspace{0.5mm}& \hspace{0.5mm} 0 \hspace{0.5mm}  & \hspace{0.5mm} 0   \hspace{0.5mm} & \hspace{0.5mm} 0  \hspace{0.5mm} & \hspace{0.5mm} 0   \hspace{0.5mm} & \hspace{0.5mm}0  \hspace{0.5mm} & \hspace{0.5mm} 0  \hspace{0.5mm} & \hspace{0.5mm} 0   \hspace{0.5mm} & \hspace{0.5mm} 0   \hspace{0.5mm} & \hspace{0.5mm}0     \hspace{0.5mm} & \hspace{0.5mm} 0  \hspace{0.5mm} & \hspace{0.5mm} 1 \hspace{0.5mm}& \hspace{0.5mm}0   \hspace{0.5mm} & \hspace{0.5mm} 0  \hspace{0.5mm} & \hspace{0.5mm} 0  \hspace{0.5mm} \\

\hspace{0.5mm} 0   \hspace{0.5mm}& \hspace{0.5mm} 0  \hspace{0.5mm}  & \hspace{0.5mm} 0  \hspace{0.5mm} & \hspace{0.5mm} 0 \hspace{0.5mm} & \hspace{0.5mm} 0   \hspace{0.5mm} & \hspace{0.5mm}0  \hspace{0.5mm} & \hspace{0.5mm}0 \hspace{0.5mm} & \hspace{0.5mm}0   \hspace{0.5mm} & \hspace{0.5mm} 0  \hspace{0.5mm} & \hspace{0.5mm} 0    \hspace{0.5mm} & \hspace{0.5mm} 0  \hspace{0.5mm} & \hspace{0.5mm} 0  \hspace{0.5mm}& \hspace{0.5mm} 0  \hspace{0.5mm} & \hspace{0.5mm} 0   \hspace{0.5mm} & \hspace{0.5mm}0   \hspace{0.5mm} 
\label{EQP} 

\end{array} \right) \ .   \ee
We determined the coefficients of the duality in terms of $k_3$ by demanding that, together with a subsequent rank  1 T-duality \footnote{Where the symmetric Jordan cross product is defined as the matrix of minors, \ie  ${\bf z}\times{\bf  z} = \det {\bf z}\, {\bf z}^{-1}$, or more generally from $\mbox{tr} \, {\bf z} ({\bf z}\times {\bf z}) = 3 \det {\bf z}$.}
\begin{equation}
  {\bf z} \rightarrow \frac{ {\bf z} - 2 \, {\bf k} \times ( {\bf z}\times {\bf z})}{1- \mbox{tr}\,  {\bf k} \, {\bf z}} \ , 
\end{equation}
with 
\begin{equation}
 {\bf k}\hspace{0.9mm} \hat{=} \hspace{0.9mm}\left(\begin{array}{ccc} \ 0 \ & \  0  \ & \ 0 \ \\   \ 0 \ & \  0  \ & \ 0 \ \\   \ 0 \ & \  0  \ & \ k_3 \ \end{array}\right)   \ , 
\end{equation}
one finds that the system includes the JMaRT solution. 

The system obtained from this duality, written in terms of the functions defining the floating brane ansatz as in \cite{Bossard:2014yta}, is very complicated and does not have any manifest symmetry. We have therefore used an appropriate change of variables to obtain a reasonable ansatz for the solution. The explicit change of variables is not very illuminating and we will not display it in this paper. It turns out that the explicit dependence in $k_3$ can be reabsorbed in a redefinition of the functions defining the ansatz, so that the system does not depend on it. In the next subsection we will describe the explicit final form of the ansatz within the STU model, which is obtained from the above by restricting all three by three symmetric matrices to be diagonal. The generalisation to an arbitrary model in the classical Jordan family is straightforward \cite{Bossard:2014yta}, but given the complexity of the system we prefer to restrict ourselves to the STU model. 

\subsection{The floating JMaRT ansatz}
\label{sec:sys-ans}
In this section we describe the ansatz for solutions to $\cN=1$ supergravity coupled to two vector multiplets in five dimensions. This theory is determined by the cubic norm 
\begin{equation}\label{eq:5d-prep}
 \mathcal{N}[t] = t^1 t^2 t^3 = \frac{1}{2} t^3 \eta_{ab} t^a t^b  \, . 
\end{equation}
However, the system is not manifestly symmetric under permutation of the three indices, and we shall choose the third direction as being special, and will often make use of the off diagonal metric  $\eta_{ab}$,
\begin{equation}\label{eq:STU-eta}
\eta_{ab}  = \left(\begin{array}{cc} 0\ &\ 1\\1\ &\ 0\end{array}\right) \,,
\end{equation}
and its inverse, $\eta^{ab}$, with $a=1,2$. 

The system is based on a solution to the Euclidean Maxwell--Einstein equations, so we will first describe this subsystem in terms of the Ernst potentials. Because the corresponding Euclidean metric will not appear as a component of our five-dimensional metric, we will rather describe this system as an $SL(3)/GL(2)$ non-linear sigma model coupled to Euclidean gravity in three dimensions. It is determined in terms of four Ernst potentials $\cE_\pm$ and $\Ppm$, which satisfy the equations 
\begin{eqnarray} \label{eq:EK-eoms}
\scal{ \cE_+ + \cE_- + \Pp  \Pm  } \Delta \cE_\pm   &=& 2 ( \nabla \cE_\pm +  \Pmp \nabla \Ppm ) \nabla \cE_\pm \  , \CR
 \scal{ \cE_+ + \cE_- + \Pp  \Pm  } \Delta \Ppm   &=& 2 ( \nabla \cE_\pm +  \Pmp \nabla \Ppm ) \nabla \Ppm \ , 
\end{eqnarray}
and determine the three-dimensional Riemannian metric $\gamma_{ij}$ through 
\begin{equation}\label{eq:R-base}
R(\gamma)_{ij} =  \frac{ ( \partial_{(i} \cE_+ + \Pm  \partial_{(i} \Pp  ) (  \partial_{j)} \cE_- + \Pp   \partial_{j)}  
\Pm  )}{( \cE_+  + \cE_- + \Pp  \Pm  )^2} - \frac{\partial_{(i} \Pp    \partial_{j)} \Pm  }{ \cE_+  + \cE_- + \Pp  \Pm  }    \  . 
\end{equation}
The corresponding four-dimensional metric is determined by the potential $V$ and the vector $\sigma$, defined as
\be \label{DefVw}
V^{-1} =\, \cE_+ + \cE_- + \Pp  \Pm \ , \qquad
\star d\sigma = V^2\, (d \cE_+ - d \cE_- + \Pm  d \Pp  - \Pp  d \Pm )\  . 
\ee
Although this metric does not define the Euclidean base of our Minkowski signature five-dimensional metric, it is convenient to use the function $V$ and the vector $\sigma$ to write explicitly the ansatz. 

The five-dimensional metric is defined as
\begin{equation}\label{eq:metric5-gen}
 ds_5^2= - \frac{W}{(H_1 H_2 H_3 )^{2/3}}(dt + k)^2 
  + (H_1 H_2 H_3 )^{1/3} \left( \frac1{W}\,(d\psi + w^0)^2 + \gamma_{ij} dx^i dx^j \right)\,,
\end{equation}
where the four-dimensional vector $k$ is defined as
\begin{gather}\label{eq:5t4k}
 k = \omega + \frac{\mu}{W} (d\psi + w^0)\ , 
\end{gather}
so that the ansatz depends on the five functions $W$, $H_I$ and $\mu$, and the two vectors $\omega$ and $w^0$ on the three-dimensional base space. It is understood everywhere that $I$ runs from 1 to 3, whereas $a$ runs from 1 to 2. Within the floating brane ansatz, we had $W = V^2$ and $w^0 = \sigma$, such that the Euclidean base was the solution to the Maxwell--Einstein equations, but this is not the case in this system. 

These functions are determined in terms of the four Ernst potentials and the six additional functions $L^I$ and $K_I$ as
\begin{align}\label{eq:scal-facts}
 W = &\, \frac1{16}\,(L^3)^2 -\frac1{4}\,V\,K_1 K_2 K_3\, \Pm \,,
 \nonumber\\
 H_a = &\, \frac14\, \eta_{ab}L^b ( L^3 - V\, \Pm  K_c L^c) + \frac14\,( V \, \Pm  L^1 L^2 -K_3) K_a \,,
 \nonumber\\
 H_3 = &\, \frac14\,V \,(\cE_- + \Pp  \Pm ) \left( (1 - V\,\cE_+)\,K_1 K_2 - \cE_+ L^3 \right) + \frac14\,V \,\cE_+^2 \Pm  K_3 \,,
\end{align}
and 
\begin{multline} \label{muequation}
\mu = \, -W\,\Pp  -\frac1{16}\,\left( 2\,(1 - V\,\cE_+)\, K_1 K_2 - \cE_+ L^3 \right)\, (K_3 + V\,\Pm  L^1 L^2 ) 
\\
       -\frac1{16}\,V\, \left(2\, \cE_+ \Pm  K_3 - (\cE_- + \Pp  \Pm )\,L^3 \right) \, K_a L^a \,,
\end{multline} 
whereas the vector fields are determined by the first order equations 
\begin{align} \label{w0Eq} 
\star d w^0 =&\, \frac14\,d L^3 - \frac12\,V\,\Pm  K_a d L^a - \frac12\,V\,K_3\,\Pm  d \cE_+ 
\CR  
&\,+ \frac12\,V\,(L^3+V\,K1 K2)\,\Pm  d \Pp   
   + \frac14\,K_1 K_2\, d V + \frac14\,K_1 K_2 \star d \sigma   \, , 
 \end{align}
and
\begin{multline}\label{eq:5d-k}
 4\,\star d \omega = \, d\scal{ \Pp  L^3 -  \cE_+ K_3}  + V\, \cE_-( K_a d L^a - L^a d K_a )
\\  +2\, V\,\cE_- \left( K_3\, d\cE_+ - ( L^3 + V\, K_1 K_2 ) d \Pp \right) 
 -V\,\Pp  \Pm  d( K_a L^a) + V\, K_a L^a \,d \cE_+ 
\\  + V\,\cE_+ \Pm   d( L^1 L^2) +V\,(\Pm  d\cE_+ - \cE_+ d \Pm ) L^1 L^2 
\\  -2\,V^2 \Pp  ( d\cE_- + \Pp  d \Pm ) K_1 K_2
 -  \cE_+ (K_a L^a + \Pm  L^1 L^2 ) \star d \sigma\, ,
\end{multline} 
where we have used everywhere the function $V$ and the vector $\sigma$ defined in \eqref{DefVw}. As the reader easily realises, the ansatz is considerably more complicated than the floating brane ansatz. Nonetheless, it turns out that the defining functions $\cE_\pm$, $\Ppm$,  $L^I$ and $K_I$ take a rather simple form for explicit solutions and one should not get discouraged prematurely by this apparent complexity. 

The two real scalar fields are defined by the three components of unit product 
\begin{equation}\label{eq:5dscal}
 t^I = \frac{(H_1 H_2 H_3 )^{1/3}}{H_I}\,.
\end{equation}
and the three vector fields decompose as usual on the three-dimensional base as
\begin{align}\label{eq:5d-gauge}
A^I = A^I_t\, (dt + k) + A^I_\psi\,(d\psi + w^0) +  w^I\,,
\end{align}
where the scalar functions $A^I_t$ and $A^I_\psi$ are determined in terms of the functions defining the system as
\begin{align}\label{eq:5dzeta}
 A^1_t = &\, \frac1{4\,H_1}\,\left( L^3 - 2\, V\, \Pm \,K_2 L^2 \right)\,,
 \CR
 A^2_t = &\, \frac1{4\,H_2}\,\left( L^3 - 2\, V\, \Pm \,K_1 L^1 \right)\,,
 \CR
 A^3_t= &\, \frac{1}{4\,H_3}\,\left((2\,V\,\cE_+ - 1) L^3 + 2\,(V\,\cE_+ - 1)\,V\,K_1 K_2 +2\,V\,K_3 \Pm \cE_+   \right)\,. 
\end{align}
and
\begin{align}\label{eq:5da}
 A^a_\psi = &\, 
 -\frac{V}{8\,W}\,\biggl(   \eta^{ab} K_b \left( (\cE_- + \Pp  \Pm )\, L^3 - 2 \,\cE_+ \Pm K_3  \right) 
 \biggr.
 \CR
& \qquad \qquad \quad \biggl.
      -  \Pm  L^a \left( 2\,K_1 K_2 (1 - V \cE_+) - \cE_+ L^3 \right)  \biggr) \,,
 \CR
 A^3_\psi= &\, 
  -\frac{1}{8\,W}\,
  \left(L^3 + 2\,V\,(K_1 K_2 - \Pm  K_a L^a) \right)\left( K_3 - V ( K_a L^a - \Pm  L^1 L^2 ) \right) 
 \CR
& \qquad \qquad \quad 
      -\frac{V^2}{4\,W}\ K_a L^a  (K_1 - \Pm  L^2)(K_2 - \Pm  L^1) \, , 
\end{align}
while the three-dimensional vector fields $w^I$ are determined by the first order equations 
\begin{eqnarray}  \label{eq:alm-NE-el}
\star d w^3 &=& \frac12\, d K_3 - \frac12\, V\, d (\Pm  L^1 L^2 ) 
- \frac12\, V\, \bigl( L^a d (K_a + \Pm  L_a) - (K_a + \Pm  L_a)\,d L^a \bigr) \,
\CR
&& +  V\,K_3\,d \cE_+ - V\,(L^3 + V\, K_1 K_2)\, d \Pp  + 2\,V\,L^1 L^2 \, d \Pm 
  + \frac12\, ( L^a  K_a + \Pm  L^1 L^2) \star d \sigma \, , \CR
  \star d w^a &=& \frac12\,d \bigl( \eta^{ab}K_b - \cE_+ V\, (\eta^{ab}K_b + \Pm  L^a) \bigr)  + \cE_+ V\, L^a d \Pm  
\CR
 &&\hspace{20mm}  - \cE_+ V^2\,(\eta^{ab}K_b + \Pm  L^a) (d \cE_- + \Pp  d \Pm  ) \   .
 \end{eqnarray}
This ansatz is rather complicated and one may wonder if it is possible to simplify it. However, we note that the ansatz obtained by the action of the duality transformation on the floating brane ansatz was considerably more complicated, and we have not been able to simplify it any further. 

We now briefly summarise the conditions for obtaining solutions. The equations of motion satisfied by the Ernst potentials $\cE_\pm,\, \Ppm$ and the Euclidean three-dimensional base metric are displayed in \eqref{eq:EK-eoms}-\eqref{eq:R-base}. The four functions $L^a$ and $K_a$ solve linear equations defined by the Bianchi identities for the two vectors $w^a$ in \eqref{eq:alm-NE-el}, and the two vectors $v_a$ as
\begin{align}\label{eq:alm-NE-mag}
 \star d v_a = &\, -2\,\eta_{ab}L^b \, V\,d \cE_- - V\,d\left( \Pp  K_a - (\cE_+ + \cE_-) \eta_{ab}L^b \right) 
 \CR
  &\,  + \left( \Pp  K_a - (\cE_+ + \cE_-) \eta_{ab}L^b \right) \star d \sigma 
\ .  
\end{align}
These two additional vectors do not appear explicitly in our ansatz, and would in fact appear in the ansatz for the 2-form potentials $B_a$ dual to the 1-forms $A^a$. The two remaining functions $K_3$ and $L^3$ solve linear equations with sources quadratic in the functions $L^a,\, K_a$, which are defined as the Bianchi identities for $w^0$, $\omega$ and $w^3$ following respectively from  \eqref{w0Eq}, \eqref{eq:5d-k} and \eqref{eq:alm-NE-el}. 
These equations altogether imply that all the five-dimensional field equations are satisfied. It would be rather cumbersome to check this explicitly in five dimensions,  and we rather rely on the analysis of the previous subsection to establish that the corresponding three-dimensional field equations \eqref{eq:3Deoms} are all satisfied.

To conclude this section we give the relevant formulae to lift the ansatz to six dimensions, in view of the fact that the system describes solutions to $\cN=(1,0)$ supergravity coupled to tensor multiplets. The metric in Einstein frame takes the form 
\be ds^2 = \frac{H_3}{\sqrt{H_1 H_2}} ( dy + A^3)^2 - \frac{W}{H_3 \sqrt{H_1 H_2}} ( dt + k )^2 + \sqrt{H_1 H_2 } \Scal{ \frac{1}{W} ( d\psi + w^0 )^2 + \gamma_{ij} dx^i dx^j} \ ,\ee
where $A^3$ is the vector field defined in \eqref{eq:5d-gauge}. It is natural to choose the vector $A^3$ as the Kaluza--Klein vector in six dimensions, because the system then generalises straightforwardly to any model of the classical Jordan family descending from  $\cN=(1,0)$ supergravity coupled to $n$ tensor multiplets in six dimensions. For one tensor multiplet, the theory includes only one scalar field that can be identified with the dilaton of the D1-D5 system, and which is given by 
\be e^{2\phi} = \frac{H_1}{H_2} \ . \ee
The six dimensional Ramond-Ramond 2-form field can then be defined as 
\be B = ( dy + A^3 ) \wedge A^1 + B_2 \ , \ee
with $A^I$ defined as in \eqref{eq:5d-gauge}, and where the 2-form field $B_2$ is dual to the vector field $A^2$ in five dimensions, \ie   
\begin{equation}\label{eq:B2-dual}
 dB_2  =  \Bigl( \frac{H_2^{\; 2}}{H_1 H_3}\Bigr)^{\frac{2}{3}} \star_5 F^2  - A^1 \wedge F^3 \ .
\end{equation}
It follows that the corresponding three-form field strength becomes 
\be H \equiv dB = - F^1 \wedge  ( dy + A^3 ) +\Bigl( \frac{H_2^{\; 2}}{H_1 H_3}\Bigr)^{\frac{2}{3}} \star_5 F^2  \ . \ee
The explicit ansatz for the 2-form $B_2$ can be found by using the expressions given above for the components of the gauge fields \eqref{eq:5d-gauge} in five dimensions. In particular, equation \eqref{eq:B2-dual} implies that $B_2$ depends explicitly on the vector field $v_2$ defined in \eqref{eq:alm-NE-mag}, as well as the vector field dual to the four-dimensional axion field. The equations of motion for the Ramond-Ramond 2-form field are then equivalent to a first order equation, relating $B$ to the dual two-form field $\tilde{B}$ in six dimensions, 
\be e^{\phi} \star_6 H +e^{-\phi} \tilde{H} = 0  \ . \ee
Here, $\tilde{H}\equiv d \tilde{B}$ and the dual two-form itself is defined in exactly the same way as $B$, by 
\be \tilde{B} =   ( dy + A^3 )  \wedge A^2 + B_1 \ .  \ee
With these formulae, one may now write a purely six-dimensional ansatz for the solutions to the floating JMaRT system.

\section{Explicit solutions}
\label{sec:ex-sols}

We now turn to an investigation of solutions to the floating JMaRT system described in the previous section.
Given the relative complexity of the ansatz, we will proceed in steps, first presenting a class of multi-centre
Einstein--Maxwell instantons generalising the ones found in \cite{Bossard:2014yta} in section \ref{sec:ein-max}, which we use in section \ref{sec:gen-sol} to built a corresponding class of explicit solutions to the floating JMaRT system that describe
a single non-extremal centre interacting with an arbitrary number of extremal centres. We then turn to the special
case of the JMaRT solution itself, which we show to be part of the general class, after fixing parameters to obtain
Minkowski asymptotics. We will not discuss the regularity of the general multi-centre solution in six dimensions in this paper, and leave this analysis for a subsequent work. 

\subsection{Multi-centre instantons}
\label{sec:ein-max}
We now discuss the multi-centre Euclidean Maxwell--Einstein instanton that we use as a base for building solutions,
following \cite{Bossard:2014yta}. 
For this purpose it is instructive to consider the BPS limit of the system, in view of the fact that we intent to construct solutions describing the interaction of a non-extremal centre with BPS centres. The latter must be such that they enter only in fields relevant for an extremal truncation, which we choοse to preserve supersymmetry. Consider the truncation where $\cE_-$ and $\Pm$ are trivial, as
\begin{equation}\label{BPStrunc}
\cE_+ = \frac{1}{\cH^0} - 1 \,, \qquad \cE_- = 1 \,, \qquad 
\Pp = \frac{\cH^3}{\cH^0}\,, \qquad \Pm=0\,,
\end{equation}
where $\cH^0$ and $\cH^3$ are arbitrary harmonic functions on $\mathbb{R}^3$ and $h$ is a constant. These are a solution to \eqref{eq:EK-eoms} and lead to a flat metric $\gamma_{ij}$ in \eqref{eq:R-base}, so that one may straightforwardly solve the remaining equations \eqref{w0Eq}, \eqref{eq:alm-NE-el} and \eqref{eq:alm-NE-mag} in terms of harmonic functions on flat space. We find the following expressions 
\begin{gather}
K_a = \eta_{ab}\frac{\cH^b}{\cH^0} \,, \qquad 
L^a = \eta^{ab} \cH_a + \frac{\cH^3 \cH^a}{\cH^0}\,, \qquad 
\CR
L^3 = \cH_3 \,, \qquad 
K_3 = \cH^0 \cH_0 + \cH^3 \cH_3  \,, \qquad 
\label{BPSfields}
\end{gather}
where the various functions $\cH^\Lambda$ and $\cH_\Lambda$ for $\Lambda$ running from $0$ to $3$ are all harmonic and constitute a symplectic vector defining the charges carried by the solution in four dimensions. It is straightforward to verify that this truncation is equivalent to the BPS system of \cite{Denef:2000nb}, up to a gauge transformation $A^3\rightarrow A^3 + d( 2 t)$ and a symplectic transformation.

In \cite{Bossard:2014yta} it was shown that one can moreover obtain multi-centre instantons with a non-flat three-dimensional base metric, by setting instead $\cE_-,\, \Pm $ to the values they take for a given Euclidean Reissner--Nordstr\"{o}m black hole solution, and then determine $\cE_+$ and $\Pp $ in terms of one function satisfying a linear differential equation. In this section we generalise this procedure, considering a Kerr--Newman solution to start with. 

We therefore fix the fields $\cE_-$, $\Pm $ to be given by those of the standard Euclidean Kerr--Newman solution with a NUT charge, as \footnote{$m_\pm$ are determined by the mass $M$ and the NUT charge $N$ as $m_\pm = M\pm N$, whereas the electric and magnetic charges $Q,\, P$ appear in $e_\pm = \frac{1}{\sqrt{2}}( Q\pm P) $ and $c^2-a^2 = M^2 + N^2 -  Q^2 -  P^2$.}
\begin{align}\label{NonExtrMinus}
 \cE_- =&\, \frac{ r - a \cos{\theta} - m_- }{ r - a \cos{\theta} + m_- }\,,
\CR
 \Pm  =&\, \frac{ 2\,e_- }{ r - a \cos{\theta} + m_- }\,,
\end{align}
where $c$, $a$, $m_\pm$, $e_-$ are constants. For $m_-=e_-=0$,  $\cE_-$ and $\Pm $ would be trivial and  $\cE_+$ and $\Pp $ would be determined in terms of two arbitrary harmonic functions as in \eqref{BPStrunc}. Working out these
equations for $\cE_+$ and $\Pp $ in the background \eqref{NonExtrMinus}, one finds that the two equations are compatible
if and only if   
\begin{equation}\cE_+ = -1 + 2\,\left( m_{-} +\frac{c^2-a^2}{r + a \cos{\theta}} \right)^{-1} 
\scal{ m_{-} -  e_- \Pp  } \ . \end{equation}
In this case one gets a solution to the system, provided $\Pp $ satisfies
\begin{align}
\Delta \Pp  + & \, 2 \, \frac{e_-}{m_{-}-e_- \Pp } \nabla \Pp  \cdot \nabla \Pp  = 
\CR
& \frac{  2\,( c^2 -a^2)\, (r - a \cos{\theta} + m_{-}) }{(r^2-c^2 + a^2 \sin^2{\theta} ) \scal{m_{-} (r + a \cos{\theta} ) + c^2-a^2 }} \nabla (r + a \cos{\theta}) \cdot \nabla \Pp  \ . 
\label{Phim} 
\end{align}
In order to find solutions, we introduce a new variable, ${\cal H}$, such that
\begin{align}
\Pp  = \frac{m_-}{e_-} \Scal{1 - \frac{1}{{\cal H}}} \ ,
\qquad
\cE_+  = -1 + 2 \frac{r + a \cos{\theta}}{r + a \cos{\theta}+ \frac{c^2-a^2}{m_-}} \frac{1}{\cH} \ ,
\end{align}
while the scale factor takes the form
\begin{equation}V^{-1} = \cE_+ + \cE_- + \Pp  \Pm  = 
\frac{ r^2-c^2 + a^2 \sin^2{\theta} }{(r - a \cos{\theta} + m_-)(r + a \cos{\theta} + \frac{c^2-a^2}{m_-})} \frac{2}{\cH} \ . \end{equation}
With this redefinition, \eqref{Phim} becomes the following linear equation for ${\cal H}$ 
\begin{equation} \label{eq:Lapl-H}
\Delta {\cal H} =  \frac{  2\,( c^2 -a^2)\, (r - a \cos{\theta} + m_{-}) }{(r^2-c^2 + a^2 \sin^2{\theta} ) \scal{m_{-} (r + a \cos{\theta} ) + c^2-a^2 }} \nabla (r + a \cos{\theta}) \cdot \nabla {\cal H} \ . 
\end{equation}
With this ansatz, one can check that the three-dimensional energy-momentum tensor does not depend on $\cH$ and is
such that
\begin{align} 
\frac{ ( \partial_{(i} \cE_+ + \Pm  \partial_{(i} \Pp  ) (  \partial_{j)} \cE_- + \Pp   \partial_{j)}  \Pm  )}{( \cE_+  
+ \cE_- + \Pp  \Pm  )^2} - \frac{\partial_{(i} \Pp    \partial_{j)} \Pm  }{ \cE_+  + \cE_- + \Pp  \Pm  } =  &\,
\CR
\frac{c^2-a^2}{(r^2 - c^2 + a^2 \sin^2{\theta})^2 } \partial_{(i}(r + a \cos{\theta})&\,\partial_{j)} (r - a \cos{\theta}) \ , \end{align}
which is exactly that of the Euclidean Kerr metric of mass $\sqrt{c^2-a^2}$.

Therefore, the three-dimensional metric is that of Euclidean Kerr--Newman even when including additional extremal
centres. This ensures the absence of attracting forces between the centres, that would usually occur through conical singularities in the three-dimensional base metric. We will use convenient spherical coordinates $(r, \theta, \varphi)$ in which the metric takes the form
\begin{eqnarray}  \label{3Dbase} 
\gamma_{ij} dx^i dx^j &=& \biggl(  1  + \frac{ a^2 \sin^2 \theta}{r^2 - c^2  } \biggr)  dr^2 + \scal{ r^2 - c^2+a^2
\sin^2\theta} d\theta^2 +  \scal{ r^2 - c^2 } \sin^2\theta d\varphi^2
\CR
 &=& \frac{ r^2 - c^2 + a^2 \sin^2{\theta}}{r_+ r_- }( dz^2 + d\rho^2) + \rho^2 d\varphi^2 \ ,
\end{eqnarray}
which are related to Weyl coordinates through 
\begin{equation}r_\pm = \sqrt{ \rho^2 + (z\pm c)^2 } \ , \qquad 2 r = r_+ + r_- \  , \qquad 2 c \cos\theta = r_+ - r_- \ . \end{equation}
This is the three-dimensional base metric that we will use throughout the remainder of this paper.

It is now straightforward to solve \eqref{eq:Lapl-H} using the metric above in order to obtain the explicit form for $\cal{H}$.
Of course, one solution to this equation is that of the standard Kerr--Newman, which we subtract from $\cal{H}$, as
\begin{equation}
{\cal H} = \frac{ r + a \cos{\theta} + m_+ }{  r + a \cos{\theta} + \frac{c^2-a^2}{m_-} }  + H\,,
\end{equation} 
in order to connect more easily with the single-centre case. The function $H$ is simply zero for the Kerr--Newman solution, and will carry poles at additional Gibbons--Hawking centres in general. We will only consider axisymmetric solutions, in which case the additional Gibbons--Hawking centres are all on the rotation axis of the original Kerr--Newman solution. Generalising appropriately the solution of \cite{Bossard:2014yta}, one obtains the following multi-centre solution 
\begin{equation}\label{eq:H-gen}
H = h +\sum_\pA \cH_\pA  \ , 
\end{equation}
with 
\be \cH_\pA  =  \frac{2\,n_\pA}{(R_\pA-a + \tfrac{R_\pA}{|R_\pA|} m_-)\,( r + a \cos{\theta} + \frac{c^2-a^2}{m_-} )} 
\frac{ (R_\pA-a) r + (a\,R_\pA- c^2)\, \cos \theta}{\sqrt{ (R_\pA-r\cos\theta)^2 + (r^2 - c^2) \sin^2\theta}}\,, \ee
where $|R_\pA|$ are the distances of the centres from the origin in Weyl coordinates, and $n_\pA$ and $h$ are constants. Note that
$H$ has simple poles at the points $(r \cos \theta =R_\pA,\, \sin\theta=0)$, that are normalised such that the residue of $V$ is $n_\pA$.

We therefore obtain the Ernst potentials
\begin{align}\label{eq:Kerr-mod}
 \cE_+ =&\, \frac{r + a \cos{\theta} - m_+ - H\, (r + a \cos{\theta} + \frac1{m_-}(c^2 - a^2)) }
              { r + a \cos{\theta} + m_+  + H (r + a \cos{\theta} + \frac1{m_-}(c^2 - a^2) ) }\,,
\CR
 \cE_- =&\, \frac{ r - a \cos{\theta} - m_- }{ r - a \cos{\theta} + m_- }\,,
\CR
\Pp  =&\, \frac1{e_-}\,\frac{ m_- ( m_+  + H (r + a \cos{\theta} + \frac1{m_-}(c^2 - a^2))) - (c^2 - a^2) }
                           { r + a \cos{\theta} + m_+  + H (r + a \cos{\theta} + \frac1{m_-}(c^2 - a^2)) }\,,
\CR
 \Pm  =&\, \frac{ 2\,e_- }{ r - a \cos{\theta} + m_- }\,,
\end{align}
while the scale factor takes the form
\begin{align}
V^{-1}=&\, \cE_+ + \cE_- + \Pp  \Pm  
\CR
= &\,
\frac{ r^2-c^2 + a^2 \sin^2{\theta} }{(r - a \cos{\theta} + m_-)} \, \frac{2}{ r + a \cos{\theta} + m_+  + H (r + a \cos{\theta} + \frac1{m_-}(c^2 - a^2) ) } \ . 
\end{align}
The equations of motion of the non-linear sigma model reduce to the conservation of the five following currents 
\begin{align} \label{eq:currs}
 \mathcal{J}_0 = &\, \frac{dr}{r^2 - c^2 + a^2 \sin^2\theta}  
   + \frac{2\, a^2 \cos{\theta}\, (\cos{\theta}\, dr - r\, d\cos{\theta} ) }{(r^2 - c^2 + a^2 \sin^2\theta)^2}\,,
\CR
 \mathcal{J}_1 = &\, \frac{d\cos{\theta}}{r^2 - c^2 + a^2 \sin^2\theta}  
   + \frac{2\,r\, (\cos{\theta}\, dr - r\, d\cos{\theta} ) }{(r^2 - c^2 + a^2 \sin^2\theta)^2}\,,
\CR
 \mathcal{J}_2 = &\, \frac{ \cos{\theta}\, dr - r\, d\cos{\theta} }{(r^2 - c^2 + a^2 \sin^2\theta)^2}\,,
\CR
 \mathcal{J}_3= &\, d H 
 + \frac{c^2-a^2}{m_-}\,\frac{ d H\,( r-a \cos{\theta}+m_- )-H\, d ( r-a \,\cos{\theta}) }{ r^2-c^2 + a^2 \sin^2\theta }
\CR
&\, 
+2\,a\,\frac{c^2-a^2}{m_-}\,H\,(r -a \cos{\theta}+ m_- )\frac{ \cos{\theta}\, d r- r\,d \cos{\theta} }{\left(r^2-c^2 + a^2 \sin^2\theta\right)^2}\,,
\CR
 \mathcal{J}_4=&\, 
 \frac{ d H\,\scal{  r+a \cos{\theta}+\frac{c^2-a^2}{m_-} }+H\,d ( r+a \,\cos{\theta}) }{ r^2-c^2 + a^2 \sin^2\theta }
\CR
&\, 
+ 2\,a\, H \,\Scal{ r +a \cos{\theta} +\tfrac{c^2-a^2}{m_-} }\frac{ \cos{\theta}\, d r - r\,d \cos{\theta} }{\left(r^2-c^2 + a^2 \sin^2\theta\right)^2}\,,
\end{align}
which define the associated vector fields through $\mathcal{J}_\gimel = \star d W_\gimel$, as 
\begin{align} \label{eq:currs-2}
W_0 = &\,  -\frac{(r^2 - c^2)\cos\theta }{r^2 - c^2 + a^2 \sin^2 \theta}  d\varphi  \,,
\CR
W_1 = &\,\frac{r \sin^2 \theta }{r^2 - c^2 + a^2 \sin^2 \theta} d\varphi  \,,
\CR
W_2 = &\,\frac{1}{2} \frac{ \sin^2 \theta }{r^2 - c^2 + a^2 \sin^2 \theta} d\varphi  \,,
\CR
W_3= &\, H \frac{c^2 - a^2}{m_-} \biggl( \cos \theta + 
      a \sin^2\theta \frac{ r - a \cos\theta + m_-}{r^2 - c^2 + a^2 \sin^2 \theta} \biggr)d\varphi    \CR
      &\qquad + \sum_\pA \cH_\pA  \biggl(  a + 
      r \cos\theta  - \frac{(R_\pA^{\; 2} - c^2)( r + a \cos \theta + \tfrac{c^2-a^2}{m_-})}{ ( R_\pA-a) r + ( a R_\pA - c^2 ) \cos \theta } \biggr) d\varphi \ ,
 \CR
W_4=&\, H \biggl( \frac{ a \sin^2 \theta  \scal{ r + a \cos \theta + \tfrac{c^2-a^2}{m_-}}}{r^2 - c^2 + a^2 \sin^2 \theta} - \cos\theta \biggr) d\varphi  \CR
& \qquad   + \sum_\pA  \cH_\pA  \frac{ r^2 - c^2 \cos^2 \theta + \frac{c^2-a^2}{m_-} ( r - R_\pA \cos \theta ) }{(R_\pA-a)r + ( a R_\pA - c^2 ) \cos \theta } d\varphi 
\, . 
\end{align}
With these expressions, it is straightforward to obtain the explicit form of the vector fields. Given that the four-dimensional metric is not realised geometrically in the supergravity solution in the floating JMaRT system, we are not particularly interested in the explicit form of the metric. Nonetheless, all the three-dimensional vector fields of the theory are linear combinations of these five vector fields in the solution we will consider in the next section, so these expressions can be used to derive the explicit form of the five-dimensional metric and the vector fields in supergravity. We refer to \cite{Bossard:2014yta} for a more detailed discussion of the limit $a\rightarrow 0$ of these solutions.

\subsection{Solving the floating JMaRT system}
\label{sec:gen-sol}

We now present a general class of solutions to the floating JMaRT system, based on the Maxwell--Einstein background
\eqref{eq:Kerr-mod} defined in the preceding section. For simplicity, we assume the solution to be such that the conserved current $ \cV P \cV^{-1} $ in \eqref{eq:3Deoms} is a linear combination of the five independent conserved currents appearing in the Maxwell--Einstein instantons \eqref{eq:currs}. The first three of these currents do not depend on $H$ and therefore are sufficient to describe a single centre solution, while the remaining two involve $H$ nontrivially and describe the  extremal centre charges and their dipole interaction with the non-extremal background. We therefore assume the conserved current to take the form 
\be \star \cV P \cV^{-1} = \sum_\gimel {\bf Q}_{\gimel}\,  \star  \mathcal{J}_\gimel  = d \Scal{ \sum_\gimel {\bf Q}_{\gimel} \, W_\gimel} \  . \ee
This assumption basically amounts to suppose that all the extremal centres carry mutually commuting charges, such that they do not interact with each other. Note that this is a restriction we use to solve the equations of motion, which does not apply to the floating JMaRT  system in general. Using this ansatz for arbitrary charges $ {\bf Q}_{\gimel} \in \mathfrak{so}(4,4)$ the system reduces to a first order linear system that can be solved algebraically \cite{Bossard:2014yta}. We will moreover disregard non-rational function of $r\pm a\cos \theta$ and $H$, that tend to admit badly behaved logarithmic singularities.

Applying the procedure above to the vector fields $dv_2$ and $dw^1$ leads to the following expressions for the functions $K_2$ and $L^1$, and respectively to the vector fields $dv_1$ and $dw^2$  for $K_1$ and $L^2$, \ie
\begin{align}\label{eq:LK-gen}
 K_a = &\, \eta_{ab}q^b + (\cE_+ + 1) \eta_{ab}l^b\,,
\CR
 L^a = &\, p^a  - \left(\cE_+ + 1 - V^{-1} \right) \,\frac{l^a}{\Pm}\,,
\end{align}
where the constants $p^a$ and $q^a$ are related to asymptotic charges, whereas the additional integration constants $l^a$ parametrise the asymptotic value of the dilatons. Given these, one may proceed in analysing the vector fields $dw^3$, $dw^0$ and $d\omega$,
leading to the final two functions
\begin{align}\label{eq:M-gen}
 K_3 = &\, \left( \cE_+ + 1 - V^{-1}  \right)^2 \frac{V}{\Pm}\,l^1 l^2 + p^3\,V
 - \left( \frac{l^3}{\Pm} - q^3 \right) \,\left( V\,( \cE_+ + 1) - 1  \right) \,,
\CR
 L^3 = &\, - l^a K_a \,\left( V\,( \cE_+ + 1) - 1  \right)  - (p^3  \Pm + q^1 q^2)\,V 
 +\left( l^3 - q^3\,\Pm \right)\,V\,( \cE_+ + 1) \,,
 \end{align}
which completely specify the solution for all the supergravity fields. Here, $ l^3$, $p^3$ and $q^3$ are integration constants. Note that the
nontrivial multi-centre background only enters through the linear combination appearing in the functions $\cE_+$ and $V$, so that it
is possible to write the solution in a compact form, despite its relative complexity. Note that we have been able to write the entire solution in terms of the functions defining the Maxwell--Einstein instanton, so it may seem that it defines a supergravity solution for any Maxwell--Einstein solution. However, this is not the case, and this solution does not reduce to an embedding of the Maxwell--Einstein subsystem. 

In the next subsection, we are going to show that this solution includes the JMaRT solution in the absence of additional extremal centres. One therefore expects this solution to generalise the JMaRT solution with additional topological cycles. However, we will not carry out the analysis of the general solution in six dimensions necessary to establish this, and leave this problem for a forthcoming work.

\subsection{The JMaRT solution}

In this section we will consider the restriction of the solution \eqref{eq:LK-gen}-\eqref{eq:M-gen} to one centre, \ie for $H=h$ with $h$ constant, imposing moreover that it is asymptotically Minkowski at spatial infinity in five dimensions. It is important to note that the single-centre solution described by \eqref{eq:LK-gen}-\eqref{eq:M-gen} does not include any regular black hole type solution. This can be understood from the property that the Euclidean three-dimensional base metric \eqref{3Dbase}, is the analytic continuation of the standard Kerr metric with $a\rightarrow \mathrm{i}\,a$, so that one expects to describe only geometries above the regularity bound. As such, this system does not include any smooth solution in five dimensions, and must be considered as a system parametrising six-dimensional supergravity solutions. Indeed the JMaRT solution describes an over-rotating Cvetic--Youm black hole in five dimensions, and only defines a smooth solution in six dimensions  \cite{Jejjala:2005yu}. 

We will now exhibit how the JMaRT solution is embedded in our system as a five-dimensional solution, as described in \cite{Gimon:2007ps}. One finds that the constants in \eqref{eq:LK-gen}-\eqref{eq:M-gen} must be determined as
\begin{gather}
{ l^I}  = 0\,,
\qquad
H = - 1 \,,
\qquad
e_- = \frac{1 + x}{q^3}\,,
\CR
p^1 = - 1 - \frac{m_-}{2\, (1 + x)}\,q^1 q^3\,,
\qquad
p^2 = - 1 - \frac{m_-}{2\, (1 + x)}\,q^2 q^3\,,\qquad
{ p^3} = \frac{m_- }{2\, (1 + x)}\,q^1 q^2 q^3\,,
\CR
m_+  = \frac1{m_-}\,(c^2-a^2 ) - \frac{4}{q^1 q^2}\,(x-1)\,, \label{eq:par-JMaRT}
\end{gather}
where $x$ is a constant parametrising $e_-$ that is introduced for convenience. The parameter $m_-$ is also fixed as
\begin{equation}\label{eq:mm-val}
 m_- = \frac1{4}\,(c^2-a^2 ) \,q^1 q^2 - \frac{q^1 + q^2}{q^1 q^2 q^3}\,(x^2-1)  - \frac{(1 + x)^2}{(q^3)^2} \,,
\end{equation}
but we will sometimes avoid using its explicit expression for brevity. With this choice of parameters, the scale factors of
the metric take the form
\begin{align}
 W = &\, \frac{( r + x\, a\cos{\theta})^2 - (1 - x^2) (c^2 - a^2)}{(r^2 - c^2 +  a^2 \sin^2\theta)^2} \,,
 \nonumber\\
 H_I = &\, \frac{ r + x\, a\cos{\theta} + E_I}{r^2 - c^2 +  a^2 \sin^2\theta} \ ,
\end{align}
where the constants $E_I$ are given by
\begin{align}\label{eq:E-jmart}
E_I = &\, \frac{ x^2-1 }{q^{I+1} q^{I+2}} + \frac{ a^2 - c^2 }{4}\,q^{I+1} q^{I+2}\,.
\end{align}
Here, we use a convention modulo 3 for the indices, so that when $I=2$, an index $I+2$ stands for $I=1$, and respectively for the other permutations.
In order to give the gauge fields, it is convenient to introduce the conserved electric charges of the solution, which
read
\begin{align}\label{eq:Q-jmart}
Q_I = &\, 4\,\frac{ x^2-1 }{q^{I+1} q^{I+2}} - (a^2 - c^2 )\,q^{I+1} q^{I+2}\,, 
\end{align}
and satisfy the conditions
\begin{equation}
 E_I^2 = \frac1{16}\,Q_I^2 + (x^2-1)\,(a^2 - c^2)\,.
\end{equation}
In order to match to the JMaRT solution one has to impose the additional constraint  \cite{Gimon:2007ps}
\begin{align}
 \prod_I \Scal{  4 E_I + Q_I - 4\, c\, (x^2-1) } = &\,
 \CR
 16\, (x^2-1)&\, (x^2 c^2 - a^2)\, \Scal{   \sum_I Q_I + 4\, \sum_I E_I - 4\,c\,(x^2-1) }\,,
\end{align}
which arises by demanding smoothness of the six-dimensional uplift of the solution and has been used to
simplify various expressions below. Within our parametrisation, this equation simplifies to 
\be 4  \biggl( c \sum_I q^{I+1} q^{I+2} - 2 \biggr)  (x^2 - 1) = (a^2 - 
    c^2) q^1 q^2 q^3  \biggl( 2 \sum_I q^{I}-c\,  q^1 q^2 q^3   \biggr)   \ , \label{eq:JM-cons} \ee
which can be solved explicitly. As will be explained below, the JMaRT solution is only smooth if $x$ and $\frac{a}{c}$
are integers \cite{Gimon:2007ps}, so it is natural to solve this equation for $c$. However, $c$ is then a cubic root, so it is easier
to analyse the solution by solving for $x$.
We now proceed to give the various quantities specifying the solution as they follow from the general ansatz
in section \ref{sec:sys-ans}, once \eqref{eq:LK-gen}-\eqref{eq:M-gen} and the constraint \eqref{eq:JM-cons}
are used. In order to describe the metric \eqref{eq:metric5-gen} one needs the vector fields
\begin{align}\label{eq:w0-om}
 w^0 = &\, \frac{-(r^2 - c^2)\, \cos{\theta} + a\, x\, r\, \sin^2\theta}{r^2 - c^2 +  a^2 \sin^2\theta}\,d\varphi
 \,,
 \CR
 \omega = &\, - \frac{\sqrt{Q_1 Q_2 Q_3}}{8\,\sqrt{c^2 x^2 - a^2}}\,
              \frac{a\,(r - c)\, \sin^2\theta }{r^2 - c^2 +  a^2 \sin^2\theta}\,d\varphi \,,
\end{align}
as well as the function $\mu$, which reads
\begin{equation}
 \mu = \frac{\sqrt{Q_1 Q_2 Q_3}}{8\,\sqrt{c^2 x^2 - a^2}}\,
   \frac{(a\, \cos{\theta}  - c\, x)\,(r - c) - a^2 x\, \sin^2\theta }{(r^2 - c^2 +  a^2 \sin^2\theta )^2}
   \,.
\end{equation}
Note that the solution is defined in terms of rational functions only, and the square root admits the rational expression
\be \frac{\sqrt{Q_1 Q_2 Q_3}}{\sqrt{c^2 x^2 - a^2}} = \frac{ 2(a^2-c^2) \prod_I ( q^{I+1} + q^{I+2})}{ c \sum_J q^{J+1} q^{J+2} - 2} \ , \ee
where we used the solution of \eqref{eq:JM-cons} for $x$.
We then find that the expressions \eqref{eq:5dzeta} and \eqref{eq:5da} lead to the
following expression for the gauge fields in \eqref{eq:5d-gauge}
\begin{align}
 A^I = &\, \frac{Q_I}{4\,(r + x\, a\cos{\theta} + E_I) } \,( dt + \omega )  + d \lambda^I
 \CR
 &\, - \frac{\sqrt{Q_1 Q_2 Q_3}}{Q_I\,\sqrt{c^2 x^2 - a^2}}\,
 \frac{x\, (c^2 - a^2 + c\, E_I ) - a\, \cos{\theta}\, ( c\, (1 - x^2) + E_I)}{2\,(r + x\, a\cos{\theta} + E_I) }\, (d\psi + w^0) 
 \CR
 &\, + \frac{ \sqrt{Q_1 Q_2 Q_3}}{Q_I\,\sqrt{c^2 x^2 - a^2}}\,
 \frac{ (c^2 - a^2 + c\, E_I ) - r\, ( c\, (1 - x^2) + E_I)}{2\,(r^2 - c^2 +  a^2 \sin^2\theta) }
 \,  a\, \sin^2\theta\,d\varphi
 \,,
\end{align}
where the gauge fields $w^0$ and $\omega$ are as in \eqref{eq:w0-om}. The $\lambda^I$ stand for a global gauge
transformation, which we give for completeness
\begin{equation}
 \lambda^I = \left( t + \Scal{ \frac{x-1}{q^1 q^2}\,q^a+\frac{x+1}{q^3}}\,\psi , 
  \quad        t +\Scal{m_-\frac{q^3}{1+x} - \frac{x-1}{q^1} - \frac{x-1}{q^2} }\,\psi \right) \,,
\end{equation}
where we remind the reader that $m_-$ is fixed by \eqref{eq:mm-val} and \eqref{eq:JM-cons} is satisfied.

For future reference, we also record the ADM mass and angular momenta of the solution, computed in the asymptotically Minkowski space-time in five dimensions, which read
\begin{align}\label{eq:MJ-gen}
 M_{\scriptscriptstyle \rm ADM} = &\, \sum_I E_I\,, \qquad 
\CR
 J_{\varphi} = &\, \frac{a}{8}\,(a^2- c^2)\,\prod_I q^I + \frac{a}{2}\,(x^2 -1)\sum_I \frac{1}{q^I} \,, \qquad 
\CR
 J_{\psi} = &\, \frac{x}{4}\,(a^2- c^2)\,\sum_I q^I + x\, \frac{x^2 -1}{\prod_I q^I}\,.
\end{align}
Using the constraint \eqref{eq:JM-cons}, the angular momenta simplify to
\begin{align}\label{eq:MJmart}
 J_{\varphi} = \frac{a}{8}\,\sqrt{\frac{Q_1 Q_2 Q_3}{c^2 x^2 - a^2}}\,, \qquad 
 J_{\psi} = \frac{c\,x}{8}\,\sqrt{\frac{Q_1 Q_2 Q_3}{c^2 x^2 - a^2}}\,.
\end{align}

In this form it is straightforward to verify that this solution is that of JMaRT \cite{Jejjala:2005yu}, in the
alternative coordinates given in \cite{Gimon:2007ps}, if one changes the time coordinate as $t\rightarrow - t$ and imposes the identifications
\begin{align}
 a = -(m-n)\,c\,, \qquad x = - m - n  \,,
\end{align}
where $m$, $n$ must be integers for a smooth solution.
The values \eqref{eq:Q-jmart} and \eqref{eq:MJmart} for the conserved charges can be seen to violate the
over-rotating bound, as is known for the JMaRT solution. The regularity bounds in this case read \cite{Cvetic:1996kv}
\begin{align}\label{eq:reg-bounds}
 S_+ = &\, \frac{a^2 - c^2}{64}\,\frac{Q_1 Q_2 Q_3}{c^2 x^2 - a^2} - J_{\varphi}^2 >0 \,,
 \CR
 S_- = &\, \frac{c^2(x^2 -1)}{64}\,\frac{Q_1 Q_2 Q_3}{c^2 x^2 - a^2} - J_{\psi}^2 >0 \,,
\end{align}
both of which evaluate to
\begin{equation}\label{eq:bound-ev}
 S_+ = S_- = 
 -c^2 (a^2 - c^2)^2 \frac{\prod_I ( q^{I+1} + q^{I+2})^2}{ 16\, ( c\, \sum_I q^{I+1} q^{I+2} -2 )^2}\,,
\end{equation}
which is manifestly negative, as expected. Nevertheless, the solution above satisfies the BPS bounds for $a>c$ and $x>1$,
since $E_I>\frac{1}{4} Q_I$ for all $I$, so we find that
 \be  M_{\scriptscriptstyle \rm ADM}  > \frac{1}{4} \sum_I Q_I \ . \ee
Note that the extremal limit of the JMaRT solution \cite{Giusto:2004id}, with $a=c$, is not a solution of the supersymmetric truncation \eqref{BPStrunc}-\eqref{BPSfields}, but corresponds instead to a Maxwell--Einstein background in the Israel--Wilson class, for which only the $\Phi_{\scriptscriptstyle \pm}$ are nontrivial. The corresponding solvable system is then equivalent up to a duality transformation to the almost BPS system \cite{Goldstein:2008fq}, which describes non-supersymmetric extremal solutions in five dimensions.

One may consider the possibility of obtaining a smooth solution that preserves the regularity bounds, by adding more centres
to the JMaRT solution. This would amount to relaxing the requirement of a constant H imposed in the beginning of this section
and allow for a function as in \eqref{eq:H-gen}. While analysing such a solution can be complicated in general, it is
straightforward to compute its asymptotic behaviour. The asymptotic expansion of the function $H$ in  \eqref{eq:H-gen} is
\begin{equation}\label{eq:H-exp}
 H = -1 + \frac{N}{r} + \mathcal{O}(r^{-2})  \,,
\end{equation}
where $N$ is a constant depending on the parameters of the various extremal centres, that is given by
\begin{equation}
 N= \sum_\pA \frac{2\, n_\pA \, (R_\pA-a)}{(R_\pA + \frac{R_\pA}{|R_\pA|} m_- - a)} \,.
\end{equation}
Using \eqref{eq:H-exp} in the conditions for the solution to be asymptotically Minkowski, one gets that the only modified constraint is the expression of $m_+$ that gets shifted by the total charge $N$ as
\begin{equation}
 m_+  = \frac1{m_-}\,(c^2-a^2 ) - \frac{4}{q^1 q^2}\,(x-1) -N\,,
\end{equation}
whereas the other parameters in \eqref{eq:par-JMaRT}  and \eqref{eq:mm-val} are unchanged. In this case, the expressions for the $E_I$ and the charges remain as in \eqref{eq:E-jmart} and
\eqref{eq:Q-jmart}, while the angular momentum $J_{\varphi}$ in \eqref{eq:MJmart} is modified as
\begin{align}
J_{\varphi} = &\, \frac{a}{8}\,(a^2- c^2)\,\prod_I q^I + \frac{a}{2}\,(x^2 -1)\,\sum_I \frac{1}{q^I}
 +  J_N \,,
\end{align}
with $J_N$ given by 
\begin{gather}
J_N= 
 \frac{(a^2 - c^2)\,(\prod q^I)^2 + 4\,(1 + x)^2 \sum_I q^{I+1} q^{I+2} }
       {32\, q^3 (x+1)}\,\sum_\pA \frac{n_\pA }{(R_\pA + \frac{R_\pA}{|R_\pA|} m_- - a)} \,.
\end{gather}
Here, $M_{\scriptscriptstyle \rm ADM}$ and $J_{\psi}$ are given by the same expressions as in \eqref{eq:MJ-gen}. So it may appear that the regularity bound on the positivity of $S_-$ is necessarily violated, but it is only possible to rewrite $-S_-$ as a perfect square in the JMaRT solution using \eqref{eq:JM-cons}, which is a consequence of the regularity of the JMaRT solution in the interior. Therefore it does not hold for the more general case and one can in fact choose $x$ such that the regularity bound is satisfied. However, demanding regularity should also constrain $x$ in the multi-centre generalisation, and a complete analysis of the solution is required to find the possible modification of this equation due to the presence of other centres. 

Assuming that one can indeed define multi-centre globally hyperbolic smooth solutions generalising the JMaRT solution within this system in six dimensions, the asymptotic behaviour of the fields in the system does not forbid them to admit the asymptotic charges of a regular black hole in five dimensions. Moreover, this discussion only applies when the extremal centres are all aligned and carry mutually commuting charges, whereas one may consider more general situations, which might lead to even more freedom in the asymptotic behaviour of the solution.

\section{Conclusion}
\label{sec:concl}

In this paper, we introduced a new partially solvable system describing solutions to six-dimensional ungauged supergravity, which
naturally includes the JMaRT solution of \cite{Jejjala:2005yu} and allows for multi-centre generalisations. As with the
floating brane system of \cite{Bena:2009fi}, this system does not include regular non-extremal black hole solutions. However, the system does not constrain by construction the asymptotic behaviour of the solution to correspond to an over-extremal black hole, as it was the case in the floating brane system \cite{Bossard:2014yta}. It may therefore 
allow for physically acceptable microstate geometries generalising the JMaRT solution. With this goal in mind,
we have given an explicit solution generalising the JMaRT geometry by the addition of an arbitrary number of mutually non-interacting
extremal centres. While we have not studied regularity of these solutions, they are of similar structure to the ones described  in
\cite{Bossard:2014yta}, so that an appropriate tuning of parameters is expected to produce smooth solutions. One should moreover be able to define general solutions with non-mutually commuting charges at the extremal centres. Understanding the system within the inverse scattering method, could help finding the most general solution and to understand the obstruction to extend it to include regular Cvetic--Youm black holes \cite{Katsimpouri:2014ara}.

In particular, one may hope to evade the problem of a lack of spin structure appearing in solutions to the floating brane
system \cite{Gibbons:2013tqa, Bossard:2014yta}, given that the JMaRT solution is known to correspond to a well defined state in the
D1-D5 system \cite{Jejjala:2005yu}. More generally, it would be very interesting to have a group theory criterion for a given partially solvable system to admit smooth solutions associated to well defined microstates in string theory. The complete duality orbit of the floating brane system might include other inequivalent systems admitting relevant supergravity solutions.

\section*{Acknowledgement}
We thank Iosif Bena, Amitabh Virmani and Nicholas Warner for stimulating discussions.
The work of S.K. is supported by the European Research Council under
the European Union's Seventh Framework Program (FP/2007-2013)-ERC
Grant Agreement n. 307286 (XD-STRING) and by INFN.

\begin{appendix}

\section{The system in four dimensions}
\label{eq:4d-sys}

In this appendix, we discuss the floating JMaRT system after dimensional reduction to four dimensions, for
later reference. In four dimensions, $\cN=2$ STU supergravity obtained after Kaluza--Klein reduction of the
five-dimensional supergravity based on the Jordan cubic norm \eqref{eq:5d-prep} is described by the prepotential 
\begin{equation}
 F[X] = -\frac{X^1 X^2 X^3}{X^0}\,,
\end{equation}
while the generalisation of the following relations to the more general models is again straightforward.
Starting from the metric, the reduction of \eqref{eq:metric5-gen} along $\psi$ leads to the metric
\begin{equation}\label{eq:metric4-gen}
  ds_4^2= - e^{2U} (dt + \omega)^2 + e^{-2U} \gamma_{ij} dx^i dx^j  \,,
\end{equation}
with
\begin{gather} 
 e^{-4U} = W^{-1} \left(H_1 H_2 H_3 - \mu^2 \right)\,,
\end{gather}
where scale factors $W$ and $H_I$ are defined according to \eqref{eq:scal-facts}.

The scalar fields and gauge fields are given again in terms of the six functions $L^I$ and $K_I$ introduced
in section \ref{sec:sys-ans}, and the vector fields are as in \eqref{eq:alm-NE-el}-\eqref{eq:alm-NE-mag},
with the two additional electric vectors
\begin{eqnarray} \label{eq:alm-NE-mag-2}
\star d v_0 &=&  2\,V\, (\Pp  d \cE_- - \cE_- d \Pp ) + 2 \, \cE_- \Pp  \star d \sigma\,, 
\CR
\star d v_3 &=& V\, (\cE_- d \cE_+ - \cE_+ d \cE_-) - \cE_+ \cE_- \star d \sigma\ ,
\end{eqnarray}
The four-dimensional complex scalars
are given by
\begin{equation}\label{eq:4dscal}
 z^I = \frac{1}{H_I}\,( \alpha^I + \Delta + \mathrm{i}\, e^{-2U})\,,
\end{equation}
where $\Delta$ reads
\begin{equation}
 \Delta = -\frac14\, \Pp  L^3 + \frac14\, \cE_+ \,K_3 
         - \frac14\,V\, \left(\cE_+ \Pm  L^1 L^2 - \Pp \Pm  K_a L^a \right) \,,
\end{equation}
and the $\alpha^I$ are given by
\begin{align}
 \alpha^1 = &\, \frac14\, V\,\cE_- (K_1 L^1 - K_2 L^2) \,,
 \nonumber\\
 \alpha^2 = &\, \frac14\, V\,\cE_- (K_2 L^2 - K_1 L^1) \,,
 \nonumber\\
 \alpha^3 = &\, \frac14\,V \,\left( 2\,(\cE_- + \Pp  \Pm ) \Pp  ( L^3 + V\, K_1 K_2 ) - 2\,\cE_+ \Pp  \Pm \,K_3 \right.
 \nonumber\\
&\,\qquad \quad \left.   + 2\,\cE_+ \Pm  L^1 L^2 - (\cE_- + 2\,\Pp  \Pm ) K_a L^a  \right) \,.
\end{align}
Here, we remind the reader that $V$ and $\sigma$ are the scale factor and the Kaluza--Klein vector field of the
Maxwell--Einstein instanton in \eqref{DefVw}. The gauge fields and their duals are defined as 
\begin{equation}
 A^A = \zeta^A (dt + \omega) + dw^A\,,\hspace{10mm}  \tilde{A}_A = \tilde{\zeta}_A (dt + \omega) + dv_A\,,
\end{equation}
where $A$ runs from $0$ to $3$, and encompassing the five dimensional vector fields and the Kaluza--Klein vector field.
The expressions for the vector fields $w^A$ are as in \eqref{eq:alm-NE-el}, while the electric potentials $\zeta^A$ are given
by
\begin{align}\label{eq:4dzeta}
 \zeta^0 = - e^{4U} \mu \,,
 \qquad
 \zeta^I = &\, A^I_t - \Re z^I \,\zeta^0\,,
\end{align}
where we used the real part of the scalars in \eqref{eq:4dscal} both to shorten the expressions and to clarify
the connection to five dimensions, where the time-like components of the three gauge fields are given by
\eqref{eq:5dzeta} and $\mu$ and $\omega$ by \eqref{muequation} and \eqref{eq:5d-k}. As in five dimensions, one obtains a solution to the supergravity equations of motion provided the Bianchi identities for the vector defined \eqref{eq:alm-NE-el} and \eqref{eq:alm-NE-mag} are satisfied.

\end{appendix}

\bibliography{PaperG} \bibliographystyle{JHEP}

\end{document}